\documentclass[a4paper,usenatbib,times]{mn2e}
\usepackage{natbib,graphicx,amsmath,amssymb,times,fleqn,url}

\def\Pg{P_{\mathrm{g}}}
\def\rhog{\rho_{\mathrm{g}}}
\def\rhogz{\rho_{\mathrm{g},0}}
\def\rhod{\rho_{\mathrm{d}}}
\def\rhodz{\rho_{\mathrm{d},0}}
\def\vg{v_{\mathrm{g}}}
\def\vd{v_{\mathrm{d}}}
\def\cs{c_{\mathrm{s}}}
\def\cssq{c_{\mathrm{s}}^{2}}
\def\tb{t_{\mathrm{s}}}
\def\tg{t_{\mathrm{g}}}
\def\td{t_{\mathrm{d}}}
\def\tp{t_{\mathrm{p}}}

\def\ed{\epsilon_{\mathrm{d}}}
\def\et{\epsilon}

\def\rz{r_{0}}
\def\rplus{r_{+}}
\def\rpm{r_{\pm}}

\def\tp{t_{\mathrm{p}}}

\def\vg{v_{\mathrm{g}}}
\def\vd{v_{\mathrm{d}}}
\def\vgb{{\bf v}_{\mathrm{g}}}
\def\vdb{{\bf v}_{\mathrm{d}}}

\def\Pg{P}
\def\vstar{{\bf v}^{*}}
\def\vstarx{v^{*}_{x}}
\def\deltavx{\Delta v_{x}}
\def\deltavxi{\Delta v_{x,\mathrm{0}}}
\def\vgi{v_{\mathrm{g},0}}
\def\vdi{v_{\mathrm{d},0}}
\def\vgx{v_{\mathrm{g},x}}
\def\vdx{v_{\mathrm{d},x}}
\def\rhog{\rho_{\mathrm{g}}}
\def\rhod{\rho_{\mathrm{d}}}
\def\deltav{\Delta {\bf v}}

\def\rmin{r_{-1}}
\def\rz{r_{0}}
\def\rplus{r_{+1}}
\def\rpm{r_{\pm 1}}

\def\dst{\displaystyle}

\title[Analytic solutions for two-fluid dust-gas problems]{\textsc{dustybox} and \textsc{dustywave}: Two test problems for numerical simulations of two fluid astrophysical dust-gas mixtures}

\author[Laibe \& Price]{Guillaume Laibe, Daniel J. Price \\
Monash Centre for Astrophysics (MoCA) and School of Mathematical Sciences, Monash University, Clayton, Vic 3800, Australia
}
\pagerange{\pageref{firstpage}--\pageref{lastpage}} \pubyear{2011}

\begin{document}
%
%
%


\def\jnl@style{\it}
\def\aaref@jnl#1{{\jnl@style#1}}

\def\aaref@jnl#1{{\jnl@style#1}}

\def\aj{\aaref@jnl{AJ}}                   
\def\araa{\aaref@jnl{ARA\&A}}             
\def\apj{\aaref@jnl{ApJ}}                 
\def\apjl{\aaref@jnl{ApJ}}                
\def\apjs{\aaref@jnl{ApJS}}               
\def\ao{\aaref@jnl{Appl.~Opt.}}           
\def\apss{\aaref@jnl{Ap\&SS}}             
\def\aap{\aaref@jnl{A\&A}}                
\def\aapr{\aaref@jnl{A\&A~Rev.}}          
\def\aaps{\aaref@jnl{A\&AS}}              
\def\azh{\aaref@jnl{AZh}}                 
\def\baas{\aaref@jnl{BAAS}}               
\def\jrasc{\aaref@jnl{JRASC}}             
\def\memras{\aaref@jnl{MmRAS}}            
\def\mnras{\aaref@jnl{MNRAS}}             
\def\pra{\aaref@jnl{Phys.~Rev.~A}}        
\def\prb{\aaref@jnl{Phys.~Rev.~B}}        
\def\prc{\aaref@jnl{Phys.~Rev.~C}}        
\def\prd{\aaref@jnl{Phys.~Rev.~D}}        
\def\pre{\aaref@jnl{Phys.~Rev.~E}}        
\def\prl{\aaref@jnl{Phys.~Rev.~Lett.}}    
\def\pasp{\aaref@jnl{PASP}}               
\def\pasj{\aaref@jnl{PASJ}}               
\def\qjras{\aaref@jnl{QJRAS}}             
\def\skytel{\aaref@jnl{S\&T}}             
\def\solphys{\aaref@jnl{Sol.~Phys.}}      
\def\sovast{\aaref@jnl{Soviet~Ast.}}      
\def\ssr{\aaref@jnl{Space~Sci.~Rev.}}     
\def\zap{\aaref@jnl{ZAp}}                 
\def\nat{\aaref@jnl{Nature}}              
\def\iaucirc{\aaref@jnl{IAU~Circ.}}       
\def\aplett{\aaref@jnl{Astrophys.~Lett.}} 
\def\apspr{\aaref@jnl{Astrophys.~Space~Phys.~Res.}}
\def\bain{\aaref@jnl{Bull.~Astron.~Inst.~Netherlands}} 
\def\fcp{\aaref@jnl{Fund.~Cosmic~Phys.}}  
\def\gca{\aaref@jnl{Geochim.~Cosmochim.~Acta}}   
\def\grl{\aaref@jnl{Geophys.~Res.~Lett.}} 
\def\jcp{\aaref@jnl{J.~Chem.~Phys.}}      
\def\jgr{\aaref@jnl{J.~Geophys.~Res.}}    
\def\jqsrt{\aaref@jnl{J.~Quant.~Spec.~Radiat.~Transf.}}
\def\memsai{\aaref@jnl{Mem.~Soc.~Astron.~Italiana}}
\def\nphysa{\aaref@jnl{Nucl.~Phys.~A}}   
\def\physrep{\aaref@jnl{Phys.~Rep.}}   
\def\physscr{\aaref@jnl{Phys.~Scr}}   
\def\planss{\aaref@jnl{Planet.~Space~Sci.}}   
\def\procspie{\aaref@jnl{Proc.~SPIE}}   

\let\astap=\aap
\let\apjlett=\apjl
\let\apjsupp=\apjs
\let\applopt=\ao

\label{firstpage}
\bibliographystyle{mn2e}
\maketitle

\begin{abstract}
 In this paper we present the analytic solutions for two test problems involving two-fluid mixtures of dust and gas in an astrophysical context. The solutions provide a means of benchmarking numerical codes designed to simulate the non-linear dynamics of dusty gas. The first problem, \textsc{dustybox}, consists of two interpenetrating homogeneous fluids moving with relative velocity difference. We provide exact solutions to the full non-linear problem for a range of drag formulations appropriate to astrophysical fluids (i.e., various prescriptions for Epstein and Stokes drag in different regimes). The second problem, \textsc{dustywave} consists of the propagation of linear acoustic waves in a two-fluid gas-dust mixture. We provide the analytic solution for the case when the two fluids are interacting via a linear drag term. Both test problems are simple to set up in any numerical code and can be run with periodic boundary conditions. The solutions we derive are completely general with respect to both the dust-to-gas ratio and the amplitude of the drag coefficient. A stability analysis of waves in a gas-dust system is also presented, showing that sound waves in an astrophysical dust-gas mixture are linearly stable.
\end{abstract}


\section{Introduction}

 Dust -- from sub-micron-sized grains to centimetre-sized pebbles -- is involved in many astrophysical problems. In particular, it provides provide the materials from which the solid cores required for the planet formation process are built \citep[see e.g.][]{ChiangYoudin2010}. Dust grains are also the main sources of the opacities in star-forming molecular clouds, thus controlling the thermodynamics. Furthermore, observations are mostly sensitive to the dust -- rather than the gas -- emission. With the advent of the \emph{Spitzer} and \emph{Herschel} space telescopes our observational knowledge of dust at different wavelengths in young stellar and planetary objects has improved substantially. Millimetre and sub-millimetre observations will similarly be vastly improved with the arrival of \emph{ALMA} that will achieve a spatial resolution $< 0.1\arcsec$ at millimetre wavelengths \citep{ALMA2006}.

Consequently, numerical simulations of astrophysical dust-gas mixtures are essential to improve our understanding of the systems we will be able to observe. A dust-gas mixture is usually treated using a continuous two-fluid description, and a large class of numerical solvers have been developed. In an astrophysical context, two types of methods are generally adopted: grid-based codes \citep[e.g.][]{Fromang2006,PM2006,Johansen2007,Miniati2010} or particle-based Smoothed Particle Hydrodynamics (SPH) codes \citep[e.g.][]{Monaghan1997,Maddison2003,BF2005}.

However, even with a continuous description of the mixture, the equations remain too complicated to be solved analytically for most problems, which presents a major difficulty for benchmarking numerical codes. Currently, the only known analytic solution in use is the solution for two interpenetrating homogeneous flows, given, e.g., by \citet{Monaghan1995} and \citet{Miniati2010} for a linear drag regime and extended to one particular non-linear regime by \citet{PM2006}. Knowing the analytic solution even for this simple case allows a precise benchmark of the various drag prescriptions that are appropriate in different astrophysical environments (e.g., \citealt{Baines1965}). On the other hand, no usable analytic solution exists for the propagation of waves in a dust-gas mixture in a regime relevant to astrophysics, despite the rich literature on the topic in the many other areas where dust-gas mixtures are of interest (for example in aerosols, emulsions or even bubbly gases, c.f. \citealt{Marble1970,Ahuja1973,Gumerov1988,Temkin1998}). Such solutions are of great interest as 1) they constitute a demanding test for a code's accuracy since small perturbations are easily swamped by numerical noise and 2) have to be correctly simulated as they often appear in physical simulations. In the absence of such a solution, astrophysical codes \citep[e.g.][]{Youdin2007,Miniati2010,Bai2010} have generally been validated against the linear growth rates for the streaming instability \citep{Youdin2005}. Such a test problem is by definition limited to checking the growth rate of a given mode rather than validating against a full analytic solution. Another approach has been to study numerical solutions for dusty-gas shock tubes \citep{Miura1982,PM2006}, where approximate solutions can be derived \citep{Miura1982} but again no complete analytic solution exists.

 In this paper we present the full analytic solutions for two specific problems concerning two-fluid gas and dust mixtures in astrophysics. The first, \textsc{dustybox} (Sec.~\ref{sec:dustybox}), is an extension of the interpenetrating flow solutions discussed above to the main drag regimes relevant to astrophysical dusty gases (i.e., Epstein and Stokes drag at different Reynolds and Mach numbers). The second, \textsc{dustywave} (Sec.~\ref{sec:dustywave}) is the solution for linear waves in a dust-gas mixture, assuming a linear drag regime.
 
  Our aim is that these solutions will be utilised as standard tests for benchmarking numerical codes designed to simulate dusty gas in astrophysics. While it is beyond the scope of this paper to benchmark a particular code using the two tests, the solutions we have derived were developed precisely for this purpose (for a new two-fluid SPH code that we are developing) and will be used to do so in a subsequent paper.

\section{\textsc{dustybox}: Two interpenetrating fluids}
\label{sec:dustybox}
 The first test problem, \textsc{dustybox}, consists of two fluids with uniform densities $\rhog$ and $\rhod$ given a constant initial differential velocity ($\Delta {\bf v}_{\mathrm{0}} =  {\bf v}_{\rm g,0} -  {\bf v}_{\rm d,0}$). We assume that the gas pressure $\Pg$ remains constant. This test is perhaps the simplest two-fluid problem that can be set up, for example by setting up two uniform fluids in a periodic box with opposite initial velocities. Thus it is a straightforward test to perform in any numerical code. Similar tests have been considered by \citet{Monaghan1995} for a single grain with a linear drag coefficient and by \citet{PM2006} for one particular non-linear drag regime. However, the simplicity of this test means that it can be used to test the correct implementation in a numerical code of both linear and non-linear drag regimes relevant to astrophysics, for which we provide the full range of solutions.

\subsection{Equations of motion}
The simplified equations of motion are given by:
\begin{eqnarray}
\rhog \frac{\mathrm{d}\vgb}{\mathrm{d} t} & = & -K f\left(\deltav \right) \left(\vgb - \vdb \right) ,\label{eq:drag_problem1} \\
\rhod \frac{\mathrm{d}\vdb}{\mathrm{d} t} & = &  K f\left(\deltav \right) \left(\vgb - \vdb \right)  ,\label{eq:drag_problem2}
\end{eqnarray}
where  momentum is exchanged between the two phases via the drag term ($K$ being an arbitrary drag coefficient) and the function $f(\Delta {\bf v})$ specifies any non-linear functional dependence of the drag term on the differential velocity (i.e. $f=1$ in a linear drag regime). In formulating (\ref{eq:drag_problem1})-(\ref{eq:drag_problem2}) it has been assumed that the effect of the collisions between the dust particles are negligible (i.e., no dust pressure or viscosity); that the dust phase occupies a negligibly small fraction of the volume (i.e., zero volume fraction: the estimated volume fraction is $\sim 10^{-12}$ in planet forming systems); that the gas is inviscid; the two phases are in thermal equilibrium and that the only way for the two phases to exchange momentum comes from the drag term (that is, additional terms due to carried mass, Basset and Saffman forces have been neglected).

\subsection{Analytic solutions}
Defining the barycentric velocity according to
\begin{equation}
\vstar = \dst \frac{\rhog {\bf v}_{\rm g,0} + \rhod {\bf v}_{\rm d,0}}{\rhog + \rhod},
\end{equation}
and adding (\ref{eq:drag_problem1}) and (\ref{eq:drag_problem2}) shows that the solutions to this equation set are of the form:
\begin{eqnarray}
\vgb \left( t \right) & = &\dst  \vstar + \frac{\rhod}{\rhog + \rhod} \deltav \left( t \right), \label{sol_form1}\\
\vdb \left( t \right) & = &\dst  \vstar -  \frac{\rhog}{\rhog + \rhod} \deltav \left( t \right). \label{sol_form2}
\end{eqnarray}

The evolution of the differential velocity $\deltav \left( t \right)$ depends on the drag regime. If the initial velocities of the two fluids have the same direction (say $x$), Eqs. (\ref{sol_form1})--(\ref{sol_form2}) reduce to two coupled scalar equations: 
\begin{eqnarray}
\vgx \left( t \right) & = &\dst \vstarx + \frac{\rhod}{\rhog + \rhod} \deltavx \left( t \right), \\
\vdx \left( t \right) & = &\dst \vstarx -  \frac{\rhog}{\rhog + \rhod} \deltavx \left( t \right),
\end{eqnarray}
where $\deltavx$ is given by the differential equation
\begin{equation}
\frac{\mathrm{d}\deltavx}{\mathrm{d}t} = -K \left(\frac{1}{\rhog} + \frac{1}{\rhod} \right) f\left(\deltavx \right) \deltavx.
\end{equation}

\begin{table*}
\begin{center}
\begin{tabular}{lll}
\hline
Drag type & $f$ & $\deltavx \left( t \right)$  \\
\hline
Linear & $ 1 $ & $ \deltavxi e^{-K \left(\frac{1}{\rhog} + \frac{1}{\rhod} \right) t}  $ \\[3em]
Quadratic & $ \left| \deltavx \right| $ & $\dst \frac{\deltavxi}{1 + \epsilon \deltavxi K \left(\frac{1}{\rhog} + \frac{1}{\rhod} \right) t} $\\[3em]
Power-law &  $\left| \deltavx \right|^{a} $ & $\dst \frac{\deltavxi}{\left( 1 + a \left(\epsilon \deltavxi \right)^{a} K \left(\frac{1}{\rhog} + \frac{1}{\rhod} \right) t  \right)^{\frac{1}{a}}} $ \\[3em]
Third-order expansion & $ 1 + a_{3} \deltavx^{2} $, $a_{3} > 0$ & $\dst \frac{\deltavxi  e^{-K \left(\frac{1}{\rhog} + \frac{1}{\rhod} \right) t}}{ \sqrt {1 + a_{3} \deltavxi^{2} \left(1 - e^{-2K \left(\frac{1}{\rhog} + \frac{1}{\rhod} \right) t} \right) }}  $ \\ [3em]
Mixed & $ \sqrt {1 + a_{2} \deltavx^{2}} $, $a_{2} > 0$ &$\dst \frac{\epsilon}{\sqrt{a_{2}}} \sqrt{ \left( \frac{ \sinh\left( K \left(\frac{1}{\rhog} + \frac{1}{\rhod} \right) t  \right) + \sqrt{1 + a_{2} \deltavxi^{2}}  \cosh\left( K \left(\frac{1}{\rhog} + \frac{1}{\rhod} \right) t  \right) }{ \cosh\left( K \left(\frac{1}{\rhog} + \frac{1}{\rhod} \right) t  \right) + \sqrt{1 + a_{2} \deltavxi^{2}} \sinh\left( K \left(\frac{1}{\rhog} + \frac{1}{\rhod} \right) t  \right)   } \right)^{2} - 1}$ \\[3em]
\hline
\end{tabular}
\end{center}
\caption{Expressions of $\deltav \left( t \right)$ for several drag regimes $f$ in a solid-gas mixture where the two phases have initially two different velocities. The pressure and densities of the medium are constant and the volume of the dust particles is neglected. $\epsilon = +1$ if $\deltavxi > 0$ and $\epsilon = -1$ if $\deltavxi <0$.}
\label{drag_analytic}
\end{table*}
\begin{figure}
   \centering
   \includegraphics[angle=0, width=\columnwidth]{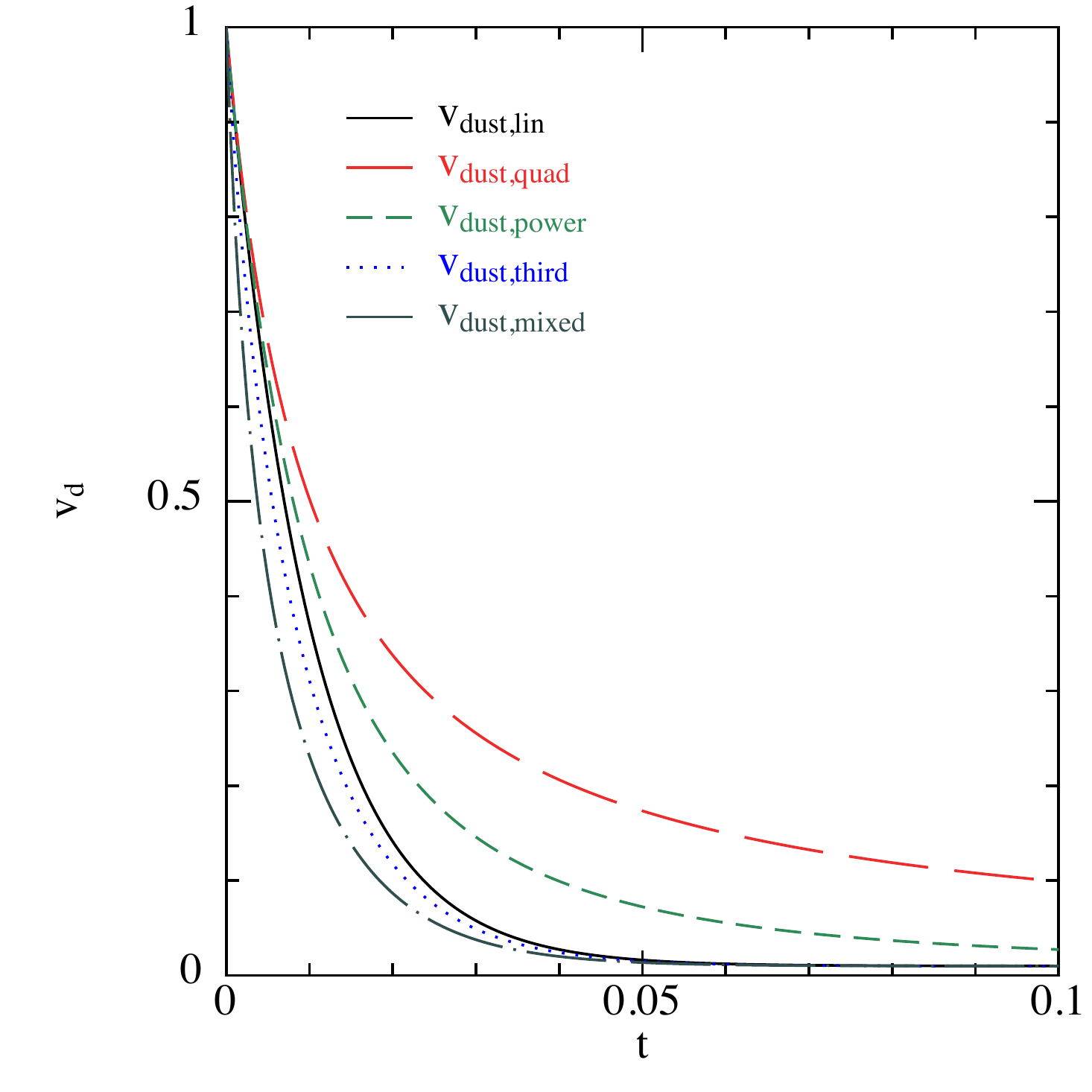} 
   \caption{Examples of the analytic solutions for the decay of the dust velocity in the \textsc{dustybox} test, assuming a dust-gas mixture with $\rhog=1$, $\rhod=0.01$, $\vdi = 1$, $\vgi = 0$ and $K=1$ for the linear, quadratic, power-law (with $a = 0.4$), third order expansion (with $a_{3} = 0.5$) and the mixed (with $a_{2} =5$) drag regimes.}
   \label{fig:dustybox}
\end{figure}

The analytic expression for $\deltavx \left( t \right)$ in five drag regimes $f(\Delta v)$ relevant to astrophysics in this particular configuration are given in Table \ref{drag_analytic}. The linear solution (top row) holds for Epstein drag at low Mach number and Stokes drag at low Reynolds number. A quadratic relation (second row) is relevant for Epstein drag at high Mach number and Stokes drag at large Reynolds numbers. Power-law drag occurs for Stokes drag at intermediate Reynolds numbers (in which case the exponent is given by $a=0.4$). The third order expansion has been proposed for Epstein drag at intermediate Mach numbers \citep{Baines1965}. The mixed drag regime (bottom row) connects the linear and quadratic regimes for Epstein drag, used recently by \citet{PM2006}.

 Finally, it should be noted that the stability of the \textsc{dustybox} problem, though likely, has only been verified numerically. Proving stability with full generality is a difficult problem due to the non-zero mean velocities for each fluid --- producing a dispersion relation that is a quadratic equation with complex coefficients. However, it can be shown that the solution is stable for particular choices of $K$, $v_{0}$, $c_{s}$ and $\rho_{0}$.
  
\subsection{\textsc{dustybox} example}
 As an example, the standard linear Epstein drag regime \citep{Baines1965} would correspond to $K=\rho_{g} c_{s} / (\rho_{int} s)$ (where $c_{s}$ is the sound speed, $\rho_{int}$ is the intrinsic density of the dust grains and $s$ is the grain size) and $f=1$. Thus, using the solution from Table \ref{drag_analytic}, we would obtain the complete expression for the velocity in each phase according to
\begin{eqnarray}
\vgx \left( t \right) & = &\dst \vstarx + \frac{\rhod}{\rhog + \rhod} \deltavxi e^{-K \left(\frac{1}{\rhog} + \frac{1}{\rhod} \right) t}  , \\
\vdx \left( t \right) & = &\dst \vstarx -  \frac{\rhog}{\rhog + \rhod}  \deltavxi e^{-K \left(\frac{1}{\rhog} + \frac{1}{\rhod} \right) t}.
\end{eqnarray}
Examples of the solutions for the decay of the dust velocity in the 5 different drag regimes for a typical astrophysical dust-gas mixture (i.e., $1\%$ dust-to-gas ratio) characterised by $\rhog=1$, $\rhod=0.01$, $\vdi = 1$, $\vgi = 0$ and assuming $K=1$ are shown in Fig.~\ref{fig:dustybox}. It may be observed, for example, that for this particular choice of parameters the quadratic and power law drag regimes (which would correspond to using a Stokes instead of an Epstein drag prescription in an accretion disc calculation) give less efficient relaxation of the dust phase to the barycentric velocity.


\section{\textsc{dustywave}: Sound waves in a dust-gas mixture}
\label{sec:dustywave}
 The second test problem, \textsc{dustywave}, consists of linear sound waves propagating in a uniform density two-fluid (dust-gas) medium with a linear drag term. Similar to the first problem, this test can easily be performed in 1, 2 or 3 dimensions in any numerical code using periodic boundary conditions. As previously, we provide analytic solutions for an arbitrary linear drag coefficient and/or dust-to-gas ratio.

 The setup consists of a sound wave propagating in the $x$-direction. We introduce the coefficient $\cs$, so that a small perturbation in the gas density $\delta \rhog$ is related to a small perturbation in the gas pressure $\delta \Pg$ by the relation $\delta \Pg = \cssq \delta \rhog$. $\cs$ is thus the sound speed of the gas phase if no dust were present. 
 
\subsection{Equations of motion}
For this system, the equations of motion are:
\begin{eqnarray}
\dst \rhog \left( \frac{\partial \vg}{\partial t} + \vg  \frac{\partial \vg}{\partial x} \right) & = &\dst -K \left(\vg - \vd \right) - \frac{\partial \Pg}{\partial x} , \label{eq:eqn_motion1} \\ [1em]
\dst \rhod \left( \frac{\partial \vd}{\partial t} + \vd  \frac{\partial \vd}{\partial x} \right) & = &\dst +K \left(\vg - \vd \right) ,\\ [1em]
\dst \frac{\partial \rhog}{\partial t} + \frac{\partial \rhog \vg}{\partial x} & = & 0, \\ [1em]
\dst \frac{\partial \rhod}{\partial t} + \frac{\partial \rhod \vd}{\partial x} & = & 0 \label{eq:eqn_motion4}.
\end{eqnarray}
The assumptions made in obtaining these equations are identical to those discussed in Sec.~\ref{sec:dustybox}, except for the additional term due to the gas pressure gradient. 

\subsection{Linear expansion}
We assume that the equilibrium velocities and densities of the fluid mixture are given by: $\vg = \vd = 0$, $\rhog = \rhogz$ and $\rhod = \rhodz$. We then consider small perturbations and perform an acoustic linear expansion of Eqs.~(\ref{eq:eqn_motion1})--(\ref{eq:eqn_motion4}):
\begin{eqnarray}
\dst \rhogz  \frac{\partial \vg}{\partial t} & = & \dst  -K \left(\vg - \vd \right) - \cssq \frac{\partial \delta \rhog}{\partial x}, \label{eq:lin1} \\ [1em]
\dst \rhodz  \frac{\partial \vd}{\partial t} & = & \dst +K \left(\vg - \vd \right) , \\ [1em]
\dst \frac{\partial \delta \rhog}{\partial t} + \rhogz \frac{\partial \vg}{\partial x} & = & 0, \\ [1em]
\dst \frac{\partial \delta \rhod}{\partial t} + \rhodz \frac{\partial \vd}{\partial x} & = & 0 . \label{eq:lin4}
\end{eqnarray}
As this system is linear, we search for solutions under the form of monochromatic plane waves. The total solution is a linear combination of those monochromatic plane waves whose coefficients are fixed by the initial conditions. The perturbations are assumed to be of the general form
\begin{eqnarray}
\vg & = & V_{\rm g} e^{i(kx - \omega t)}, \label{eq:vpert} \\
\vd & = & V_{\rm d} e^{i(kx - \omega t)}, \\
\delta\rhog & = & D_{\rm g} e^{i(kx - \omega t)}, \\
\delta\rhod & = & D_{\rm d} e^{i(kx - \omega t)}, \label{eq:dpert}
\end{eqnarray}
where in general the perturbation amplitudes $V_{\rm g}$, $V_{\rm d}$, $D_{\rm g}$ and $D_{\rm d}$ are complex quantities.

\subsection{Linear solutions}
Using (\ref{eq:vpert})--(\ref{eq:dpert}) in (\ref{eq:lin1})--(\ref{eq:lin4}), we find that the resulting system admits non-trivial solutions provided the following condition holds:
\begin{equation}
\left|
\begin{array}{cccc}
- \dst i\omega + \frac{1}{\tg} &\dst -\frac{1}{tg} &\dst +\frac{i k \cssq}{\rhogz} & 0 \\ [1em]
\dst-\frac{1}{\td} &\dst -i \omega + \frac{1}{td} & 0 & 0 \\ [1em]
i k \rhogz & 0 & -i \omega & 0 \\ [1em]
0 & i k \rhodz & 0 & -i \omega 
\end{array}
\right|
=0,
\label{det}
\end{equation}
where we have set $\tg = \dst \frac{\rhogz}{K}$ and $\td = \dst \frac{\rhodz}{K}$. This condition provides the dispersion relation of the system,
\begin{equation}
\omega ^{3} + i\omega^{2} \left(\frac{1}{\tg} + \frac{1}{\td}\right) - k^{2}\cssq \omega - i \frac{k^{2}\cssq}{\td} = 0.
\label{eq:disp_rel}
\end{equation}
This cubic equations admits three complex roots $\omega_{n=1,2,3}$ whose imaginary parts are always negative, ensuring the linear stability of the system (see Appendix \ref{sec:stab}). This implies that the full solution of the problem consists of a linear combination of three independent modes that will take the form of exponentially decaying monochromatic waves. For example, for the the gas velocity the solution will be given by
\begin{eqnarray}
\vg (x, t) & = &  e^{\omega_{1\mathrm{i}} t} \left[V_{1\mathrm{g,r}}  \cos\left(kx - \omega_{1\mathrm{r}}t \right) -  V_{1\mathrm{g,i}} \sin\left(kx - \omega_{1\mathrm{r}}t \right) \right], \nonumber \\
& + &  e^{\omega_{2\mathrm{i}} t} \left[V_{2\mathrm{g,r}}  \cos\left(kx - \omega_{2\mathrm{r}}t \right) -  V_{2\mathrm{g,i}} \sin\left(kx - \omega_{2\mathrm{r}}t \right) \right], \nonumber \\
& + &  e^{\omega_{3\mathrm{i}} t} \left[V_{3\mathrm{g,r}}  \cos\left(kx - \omega_{3\mathrm{r}}t \right) -  V_{3\mathrm{g,i}} \sin\left(kx - \omega_{3\mathrm{r}}t \right) \right]. \nonumber \\
\label{eq:Vg}
\end{eqnarray}
where the subscripts $r$ and $i$ refer to the real and imaginary parts of the complex variables $\omega_{1,2,3}$ and $V_{g,1,2,3}$. The solutions for $\vd$, $\delta\rhog$ and $\delta \rhod$ are of the same form, with the amplitudes replaced by the real and imaginary parts of $V_{\rm d}$, $D_{\rm g}$ and $D_{\rm d}$, respectively.

\subsection{Solving for the coefficients}
  Obtaining the full analytic solution thus requires two steps:
 \begin{enumerate}
 \item Solving the cubic equation (\ref{eq:disp_rel}) to determine the complex variables $\omega_{1}$, $\omega_{2}$ and $\omega_{3}$ for the 3 modes; and
 \item Solving for the 24 coefficients determining the amplitudes $V_{\rm g}$, $V_{\rm d}$, $D_{\rm g}$ and $D_{\rm d}$ for each of the 3 modes.
 \end{enumerate}
 Step i) can be achieved straightforwardly using the known analytic solutions for a cubic equation (given for completeness in Appendix~\ref{sec:cubic}). Since in general such solutions require a cubic equation with real coefficients, it is convenient to solve for the variable $\omega = -i y$, which reduces Eq.~(\ref{eq:disp_rel}) to the form
\begin{equation}
y ^{3} - y^{2} \left(\frac{1}{\tg} + \frac{1}{\td}\right) + k^{2}\cssq y - \frac{k^{2}\cssq}{\td} = 0,
\label{eq:disp_rel_simp}
\end{equation}
which has purely real coefficients, as required.

 Step ii) is less straightforward and consists of two substeps. The first substep is to constrain the amplitude coefficients using the 8 constraints given by the initial conditions (i.e., the phase and amplitude of the initial mode in the numerical simulation, which constrain both the real and imaginary parts of the initial amplitudes). Although the solution can in principle be found for any given combination of initial perturbations to $v$ and $\rho$ for the two phases, the solutions we provide assume initial conditions of the form
 \begin{eqnarray}
\vg (x, 0) & = & v_{\rm g,0} \sin(kx), \label{eq:ic1} \\
 \vd (x,0) & = & v_{\rm d,0} \sin(kx), \\
 \rhog (x, 0) & = & \rho_{\rm g,0} + \delta\rho_{\rm g,0} \sin(kx), \\
 \rhod (x,0) & = & \rho_{\rm d,0} + \delta\rho_{\rm d,0} \sin(kx), \label{eq:ic4}
 \end{eqnarray}
giving the 8 constraints
\begin{eqnarray}
V_{\rm 1g,r} + V_{\rm 2g,r} + V_{\rm 3g,r}  &= & 0, \label{eq:con1}\\
V_{\rm 1g,i} + V_{\rm 2g,i} + V_{\rm 3g,i}  &= & -v_{\rm g,0}, \\
V_{\rm 1d,r} + V_{\rm 2d,r} + V_{\rm 3d,r}  &= & 0, \\
V_{\rm 1d,i} + V_{\rm 2d,i} + V_{\rm 3d,i}  &= & -v_{\rm d,0}, \\
D_{\rm 1g,r} + D_{\rm 2g,r} + D_{\rm 3g,r}  &= & 0, \\
D_{\rm 1g,i} + D_{\rm 2g,i} + D_{\rm 3g,i}  &= & -\delta\rho_{\rm g,0}, \\
D_{\rm 1d,r} + D_{\rm 2d,r} + D_{\rm 3d,r}  &= & 0, \\
D_{\rm 1d,i} + D_{\rm 2d,i} + D_{\rm 3d,i}  &= & -\delta\rho_{\rm d,0}. \label{eq:con8}
\end{eqnarray}

\begin{figure}
   \centering
   \includegraphics[angle=0, width=\columnwidth]{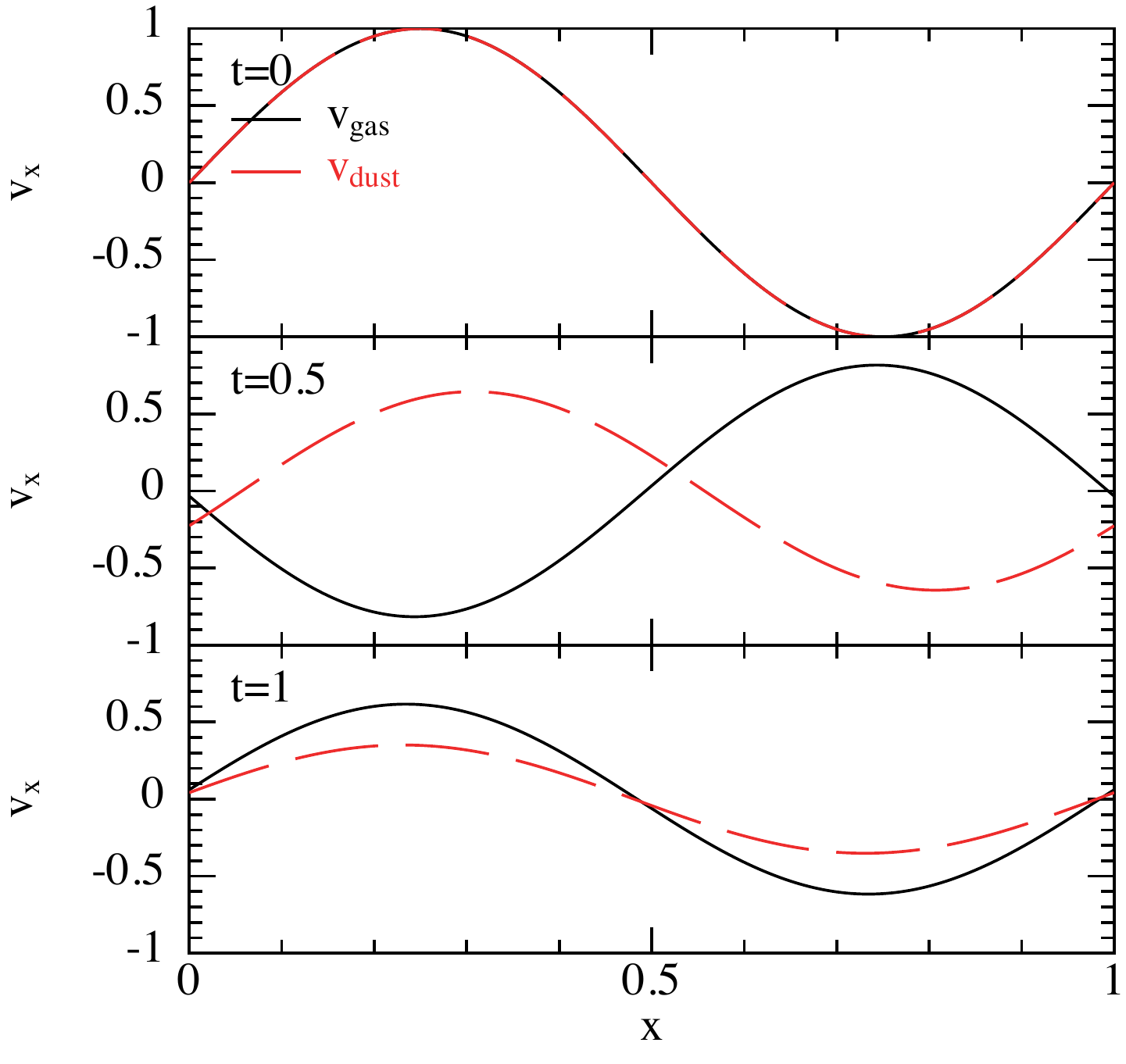} 
   \caption{Example analytic solution of the \textsc{dustywave} test showing the propagation of a sound wave in a periodic domain in a two-fluid gas-dust mixture. The panels shows the time evolution (top to bottom, time as indicated) of the velocity in the gas (solid/black) and dust (dashed/red) respectively assuming $\rhog = \rhod = 1$ (i.e., a gas-to-dust ratio of unity) and a drag coefficient $K =1$ giving a characteristic stopping time of $t_{s} = 1/2$. The solution with $K=1$ shows efficient damping of the initial perturbation in both fluids.}
   \label{fig:wave_time}
\end{figure}

\begin{figure}
   \centering
   \includegraphics[angle=0, width=\columnwidth]{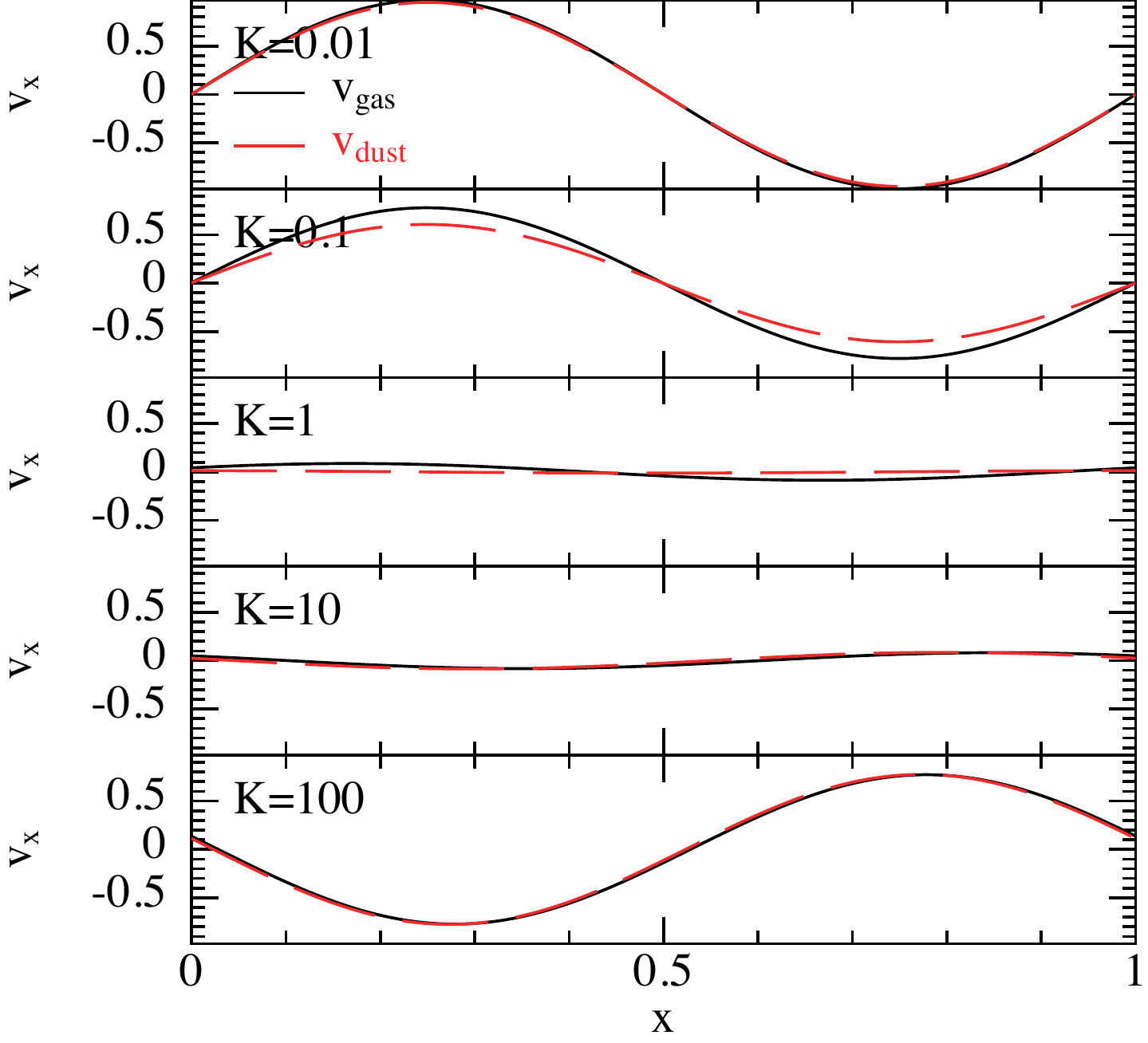} 
   \caption{Further examples of analytic solutions to the \textsc{dustywave} test, as in Fig.~\ref{fig:wave_time} but showing the solution at $t=5$ for a range of drag coefficients $K =[0.01,0.1.1.10,100]$ (top to bottom, as indicated in the legend). At low $K$ the waves are essentially decoupled in the two fluids (top panel), while an intermediate drag coefficient produces the most efficient damping (middle panels). At large $K$ (bottom panels), although the differential velocities are quickly damped the overall amplitudes decrease more slowly since the waves tend to move together.}
   \label{fig:wave_snap}
\end{figure}

The second substep is to determine the remaining 16 coefficients by substituting each of the expressions for the perturbations (Eq.~\ref{eq:Vg} and the equivalents for $v_{\rm d}$, $\delta\rhog$ and $\delta\rhod$) and the 8 constraints (\ref{eq:con1})--(\ref{eq:con8}) into the evolution equations (\ref{eq:lin1})--(\ref{eq:lin4}). The remaining analysis is straightforward but laborious, hence we perform this step using the computer algebra system \textsc{maple}. The resulting expressions for the 24 coefficients may be easily obtained in this manner, however their expressions are too lengthy to be usefully transcribed in this paper. Instead, we provide, for practical use, both the \textsc{maple} worksheet and a Fortran (90) routine\footnote{Available as supplementary files accompanying the arXiv.org version of this paper.} that evaluates the analytic expressions for the coefficients (produced via an automated translation of the \textsc{maple} output). The Fortran routine has been used to compute the example solutions shown in Figures~\ref{fig:wave_time} and \ref{fig:wave_snap}. Note that, although the initial conditions are constrained to be of the form (\ref{eq:ic1})--(\ref{eq:ic4}), the solutions provided are completely general with respect to both the amplitude of the drag coefficient and the dust-to-gas ratio. The examples we show employ a dust-to-gas ratio and drag coefficients that are typically relevant during the planet formation process. For this test it should be kept in mind that the solutions assume linearity of the wave amplitudes and do not therefore predict possible non-linear evolution of the system  --- for example the potential for mode splitting/merging or self-modulation, effects that are known to occur in multi-fluid systems.

\subsection{\textsc{dustywave} examples}
 Figure~\ref{fig:wave_time} shows a typical time evolution of the gas and dust velocities (solid/black and red/dashed lines, respectively) assuming a dust-to-gas ratio of unity (as would occur during the late stages of the planet formation process) and a drag coefficient $K=1$, giving a characteristic stopping time of $t_{s} = 1/ [ K (1/\rhog + 1/\rhod) ] = 1/2$. At $t=1$ (bottom panel) it may be observed that the differential velocity between the two phases has been efficiently damped by the mutual drag between the two fluids.

 Figure~\ref{fig:wave_snap} demonstrates how the characteristics of the solution change as the drag coefficient varies, again assuming a gas-to-dust ratio of unity, with the solution shown at $t=5$. At low drag (top panel, $K=0.01$, e.g. for dust in the interstellar medium) both the dust and gas evolve essentially independently over the timescale shown. Thus, the solution in the gas is simply a travelling wave with a sound speed close to the gas sound speed, while the dust retains its initial velocity profile. As the drag coefficient increases to unity (top three panels), the solution tends towards the efficient coupling that occurs at $K\sim 1$ (for this gas-to-dust ratio) shown in Fig.~\ref{fig:wave_time}, which represents the ``critical damping'' solution where both the gas and dust velocities relax to zero. As the damping is increased further (bottom two panels) the damping of the differential velocities occurs in a fraction of a period, implying that, although the differential velocity between the fluids is quickly damped, the removal of kinetic energy is less efficient since the two waves essentially evolve together, relaxing slowly --- but in tandem --- to zero.

\section{Conclusion}

In this paper we have provided the analytic solutions to two problems involving two-fluid astrophysical dust-gas mixtures in order to supply a practical means of benchmarking numerical simulations of dusty gas dynamics. The test problems are simple to setup for both particle and grid-based codes and can be performed using periodic boundary conditions. A summary of the setup for each problem is given in Table~\ref{tab:summary}. It may be noted that both solutions are completely general with respect to both the amplitude of the drag coefficient and the dust-to-gas ratio. The subroutines for computing the \textsc{dustywave} solution are provided as supplementary files to the version of this paper posted on arXiv.org. The \textsc{dustywave} solution has also been incorporated into the \textsc{splash}\footnote{http://users.monash.edu.au/$\sim$dprice/splash/} visualisation tool for SPH simulations \citep{Price2007}. 

\begin{table}
\begin{center}
\begin{tabular}{lp{3.5cm}l}
\hline
Test problem & Initial conditions & Boundary conds. \\
\hline
\textsc{dustybox} & $\vgb = v_{0}\hat{\bf r}$ \newline $\vdb = -v_{0}\hat{\bf r}$ \newline $\rhog = \rho_{\rm g, 0}$ \newline 
$\rhod = \rho_{\rm d,0}$ \newline
$c_{\rm s,0} = {\rm const}$
& Periodic \\
\textsc{dustywave} & $\vgb = v_{\rm g,0} \sin({\bf k}\cdot{\bf r}) \hat{\bf r}$ \newline
 $\vdb = v_{\rm d,0} \sin({\bf k}\cdot{\bf r}) \hat{\bf r}$ \newline
 $\rhog = \rho_{\rm g,0} + \delta\rho_{\rm g,0} \sin({\bf k}\cdot{\bf r})$ \newline
 $\rhod = \rho_{\rm d,0} + \delta\rho_{\rm d,0} \sin({\bf k}\cdot{\bf r})$ \newline
 $c_{\rm s,0} = {\rm const}$
& Periodic \\
\hline
\end{tabular}
\caption{Summary of the setup for each of the test problems.}
\label{tab:summary}
\end{center}
\end{table}

\section*{Acknowledgments}
We thank Matthew Bate, Ben Ayliffe and Joe Monaghan for useful discussions. We are grateful to the Australian Research Council for funding via Discovery project grant DP1094585.

\bibliography{Dust_test}

\begin{appendix}

\section{Stability of linear waves in a dust-gas mixture}
\label{sec:stab}

The roots of Eq. (\ref{eq:disp_rel_simp}) can be three real single roots or one real roots and two conjugated complex roots (plus all the degenerated cases). Noting $y =  y_{\mathrm{r}} + i y_{\mathrm{i}}$, we see e.g. from Eq. (\ref{eq:Vg}) that if $y_{\mathrm{r}}  < 0$ (i.e. $= \omega_{\mathrm{i}}>0$), the solution diverges with time and the system is unstable.

Physically, the evolution of the solid-gas mixture given by Eqs. (\ref{eq:lin1})--(\ref{eq:lin4}) is characterised by three intrinsic time scales: the time $\tp = \left( k \cs \right) ^{-1}$ required for the gas pressure to equilibrate the gas phase, the time $\tb$ (defined according to $\tb^{-1} = \tg^{-1} + \td^{-1}$) required by the drag to relax the centre of mass of the fluid (see Sec.~\ref{sec:dustybox}) and $\td$, the time required for the dust to force the gas evolution. Thus, two ratios of these independent timescales are sufficient to fully describe the evolution of the system. Defining the following dimensionless quantities:
\begin{eqnarray}
Y & \equiv & \dst y \tp , \\ [1em]
\et & \equiv & \dst \frac{\tp}{\tb} , \\ [1em]
\ed & \equiv & \dst \frac{\tp}{\td} ,\label{dimensionless}
\end{eqnarray}
giving Eq.~(\ref{eq:disp_rel_simp}) in the form
\begin{equation}
P_{0}\left(Y\right) = Y ^{3} - Y^{2} \et + Y - \ed = 0 .
\label{eq:disp_red}
\end{equation}

Determining the sign of the real part of the roots of Eq.~(\ref{eq:disp_red}) specifies whether the system is unstable or not. For this purpose, we introduce:
\begin{eqnarray}
P & = & \dst \frac{\et^{2}}{9} - \frac{1}{3} , \\ [1em]
Q & = & \dst \frac{\et ^{3}}{27} + \frac{1}{2}\left(\ed - \frac{\et}{3} \right),\\ [1em]
D & = & P^{3} - Q^{2} = \dst -\frac{\et^{3}\ed}{27} + \frac{\et^{2}}{108} + \frac{\et\ed}{6}  - \frac{\ed^{2}}{4}   - \frac{1}{27}  .
\end{eqnarray}
The roots of Eq.~(\ref{eq:disp_red}) are denoted $\rmin$, $\rz$ and $\rplus$. We have the following three cases:
\begin{itemize}
\item Case(i): $P<0$, 
\item Case (ii): $P>0$ and $D<0$, 
\item Case (iii): $P>0$ and $D>0$, 
\end{itemize}

In cases (i) and (ii) one root is real (denoted $\mu$) while the two remaining roots are the complex conjugates of each other (denoted $\alpha \pm i \beta$). Factorising Eq.~\ref{eq:disp_red} using the three roots gives the relations
\begin{eqnarray}
\mu + 2 \alpha & = & \et , \label{eq:mu} \\ [1em]
2 \mu \alpha + \left(\alpha^{2} + \beta^{2} \right) & = & 1  ,\\ [1em]
\mu \left(\alpha^{2} + \beta^{2} \right) & = & \ed. \label{eq:eps}
\end{eqnarray}
The last equation implies $\mu >0$. Combining (\ref{eq:mu})--(\ref{eq:eps}) gives an equation for $\alpha$ of the form
\begin{equation}
f\left( \alpha \right) = \alpha^{3} - \et \alpha^{2} + \frac{\alpha}{4}\left(\et^{2} + 1 \right) - \frac{ \left(\et - \ed \right)}{8} = 0 ,
\label{alpha_three}
\end{equation}
which admits only one or three positive roots provided $\et - \ed = \frac{\tp}{\tg} > 0$. Indeed, this is the case since we have $f\left( 0 \right)<0$, $f ' \left( 0 \right)>0$, $\lim\limits_{\substack{\alpha \to +\infty}} f \left( \alpha \right)$ and a positive X axis value $\alpha_{\mathrm{c}} = \dst \frac{\et}{3}$ for the centre of symmetry of the cubic function. Therefore, in cases (i) and (ii), the real part of the complex roots is positive and the system is stable.

In case (iii), the three roots are real. To determine their signs, we calculate the Sturm polynomials of $P_{0}\left(Y\right)$:
\begin{eqnarray}
P_{0}\left(Y\right) & = & \dst Y ^{3} - Y^{2} \et + Y  -  \ed , \\ [1em]
P_{1}\left(Y\right) & = & \dst 3 Y^{2} - 2 \et Y + 1,\\ [1em]
P_{2}\left(Y\right) & = & \dst - \left[ \left(\frac{2 }{3} - \frac{2 \et^{2}}{9} \right) Y +  \left(\frac{\et}{9} - \ed\right) \right], \\ [1em]
P_{3}\left(Y\right) & = & \dst \frac{243 D}{\left(\et^{2} - 3 \right)^{2}} .
\end{eqnarray}
As the three roots are real, $D>0$ and $P_{3} > 0$. We can then apply the Sturm theorem to determine the number of positive roots. Using $V \left( Y \right)$ to denote the number of consecutive sign changes in the sequence $\left[P_{0}\left(Y\right),P_{1}\left(Y\right),P_{2}\left(Y\right),P_{3}\left(Y\right) \right]$, the number of positive roots of $P_{0}\left(Y\right)$ is given by $V \left( 0 \right) - \lim\limits_{\substack{Y \to +\infty}} V \left( Y \right) $. Thus, if $ \ed -\frac{\et }{9} <0$, the three roots are positive. However, if  $ \ed -\frac{\et }{9} >0$, only one root is positive and the two remaining ones are negative. However, $\ed$ has to be smaller than the larger positive root of $D(\ed)=0$ for $D$ in order to remain positive (and thus, for the system to be unstable), which is never the case if $9\ed>\et>\sqrt{3}$. 

Such a solid-gas mixture would thus be unconditionally stable. However, if $\et$ is large enough (which means physically, that the damping due to the drag occurs faster than the equilibrium produced by the pressure) and the ratio $\dst \frac{\ed}{\et}$ is sufficiently large (i.e., a large dust-to-gas ratio and thus, an efficient forcing of the gas motion from the dust), and instability may develop when additional physical processes are involved (an example being the streaming instability that occurs in a differentially rotating flow, c.f. \citealt{Youdin2005}).

\section{Real and imaginary parts of the roots of a cubic with real coefficients}
\label{sec:cubic}

The cubic equation given by Eq.~\ref{eq:disp_rel_simp} can be solved using the known analytic solution to a cubic equation, though for this problem we require both the real and imaginary components of all three solutions. We consider the following \textit{normalised} cubic equation with respect to the variable $x$:
\begin{equation}
f \left( x \right) = x^{3} + ax^{2} + bx + c,
\label{eq:general_cubic}
\end{equation}
where $a$, $b$ and $c$ are real coefficients. We introduce the quantities $P$, $Q$ and $D$, given by:
\begin{equation}
P = \frac{a^{2} - 3b}{9},
\label{eq:defP}
\end{equation}
\begin{equation}
Q = \frac{ab}{6} - \frac{c}{2} - \frac{a^{3}}{27},
\label{eq:defQ}
\end{equation}
and
\begin{equation}
D= P^{3} - Q^{2}.
\label{eq:defD}
\end{equation}
Denoting the roots of Eq.(\ref{eq:general_cubic}) by $\rmin$, $\rz$ and $\rplus$, the solutions can be obtained by considering the following three cases:

\begin{enumerate}
\item[Case i)] $P<0$,
\begin{eqnarray}
\rz & = & \dst \frac{-a}{3} + 2\sqrt{-P} \, \sinh\left( t \right) , \\ [1em]
\rpm & = &  \dst \frac{-a}{3} - \sqrt{-P} \, \sinh\left( t \right) \pm i \sqrt{-3P} \,  \mathrm{cosh}\left( t \right) .
\label{case_i}
\end{eqnarray}
where
\begin{equation}
t = \frac{1}{3}\, \mathrm{arcsinh}\left( \frac{Q}{\sqrt{\left(-P \right)^{3}}} \right) .
\label{eq:deft}
\end{equation}

\item[Case ii)] $P>0$ and $D<0$, 
\begin{eqnarray}
\rz & = & \dst \frac{-a}{3} + u + v , \\ [1em]
\rpm & = & \dst   \frac{-a}{3} - \frac{u + v}{2} \pm i \sqrt{3}\frac{u - v}{2} .
\end{eqnarray}
where:
\begin{eqnarray}
u & = & \dst \sqrt[3]{Q - \sqrt{-D}} , \\  [1em] \label{eq:defu}
v & = & \dst  \sqrt[3]{Q + \sqrt{-D}}  .\label{eq:defv}
\end{eqnarray}

\item[Case iii)] $P>0$ and $D>0$, 
\begin{equation}
\dst r_{n} = \frac{-a}{3} + 2 \sqrt{P} \mathrm{cos}\left(\dst \frac{2 \pi n + \mathrm{arccos}\left(\frac{Q}{\sqrt{P^{3}}} \right)}{3} \right), 
\end{equation}
with $n = 0, \pm 1$.
\end{enumerate}

\end{appendix}

\label{lastpage}
\end{document}